# Evidence for a Critical Behavior in 4D Pure Compact QED

J. Jersák[a], C.B. Lang[b] and T. Neuhaus[c]

[a]Institut für Theoretische Physik E, RWTH Aachen and HLRZ Jülich, FRG

[b]Institut für Theoretische Physik, Universität Graz, Austria

[c]FB8 Physik BUGH Wuppertal, FRG

We present evidence about a critical behavior of 4D compact QED (CQED) pure gauge theory. Regularizing the theory on lattices homotopic to a sphere, we present evidence for a critical, i.e. second order like behavior at the deconfinement phase transition for certain values of the coupling parameter $\gamma$.

## 1. Introduction

Four dimensional CQED pure gauge theory is a prototype model of gauge interactions of abelian nature. It has a deconfinement phase transition at values of the gauge coupling constant of order 1, separating confinement and Coulomb phases of the theory. The order of the deconfinement phase transition in the $\beta - \gamma$ plane of couplings

$$S = -\sum_P [\beta \cos(U_P) + \gamma \cos(2U_P)] \qquad (1)$$

has been studied for long time. In the early times of LGT the transition on the Wilson line ($\gamma = 0$) was thought to be continuous. Later the tricritical point (TCP) was suggested [1] at a slightly negative value of $\gamma_{TCP} = -0.11$, predicting that the deconfinement phase transition on the Wilson line is of weak first order.

The physically interesting question, whether there really exists a critical point in 4D CQED and if so, what are its properties, has been left unanswered throughout the years. One can approach this question by a determination of the correlation length critical exponent $\nu$. A value of $\nu = \frac{1}{2}$ can be expected assuming a mean field behavior. A value of $\nu = \frac{1}{D}$ would correspond to a discontinuous phase transition. Other and different values of $\nu$ would be indicative of an "unexpected new" behavior. The existence of such a topological disorder driven fixed point could have implications for the continuum limit of 4D field theories containing compact $U(1)$ gauge degrees of freedom, especially when fermions are included.

Although a critical transition is expected to exist at least at sufficiently large negative $\gamma$-values [1,2], this was never confirmed in the stochastic evaluation of the path integral. In fig. 1 we display the probability distribution of the normalized plaquette expectation value evaluated close to the deconfinement phase transition at a $\gamma$-value of $\gamma = -0.2$ on a $8^4$ lattice with p.b.c. As can be seen, double peak structure is present and actually can be observed up to quite sizable negative values of $\gamma$ [1,3]. The appearance of such signals may cause doubts about the order of the transition and excludes a meaningful finite size scaling (FSS) analysis of the critical behavior of CQED.

We have argued [4], that the disturbing effects may be due to an inappropriate choice of the boundary conditions of the system. Therefore there is need for a improved lattice regularization of CQED. Almost all previous numerical simulations of the CQED path integral have been performed on a torus with p.b.c.. We proposed to use lattices topologically homotopic to a sphere. On such a lattice any monopole loop, closed on the dual lattice, can be contracted to a point, while on a torus this cannot be achieved for all loops. The actual construction of the lattice was performed by considering the manifold given by the surface of a 5-dimensional hypercube with linear extent $L$, denoted $SH(L)$ here. Simulations on $SH(L)$ on the Wilson line demonstrated somewhat sur-prisingly the absence of metastability signals.

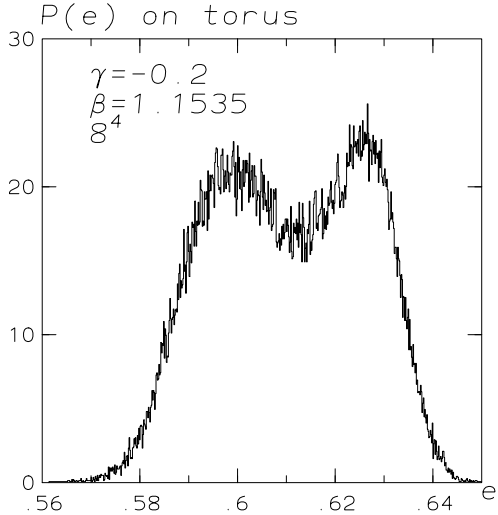

Figure 1. $P(e)$ on torus at $\gamma = -0.2$.

However, in that study the plaquette terms of the lattice action of (1) were discretized in the most naive way. Plaquettes in $SH(L)$ were assigned equal weight to the action function. Thus, although the global topological properties of a sphere were implemented, effects related to the curvature of the sphere were not. In that geometry the curvature is distributed inhomogeneuosly and therefore space-time symmetries present for the torus are violated. These effects have been studied and no consistent FSS at criticality has been observed. Instead the data result into an upper bound on the discontinuity in the plaquette operator on the Wilson line, consistent with the earlier extrapolation in the vicinity of the TCP[4].

## 2. CQED on Almost Perfect Spheres

The construction of an improved lattice action of CQED on a sphere is guided by the assumption, that certain non-universal features of the lattice discretization should become irrelevant in the vicinity of a field LGT fixed point. Therefore we may as well use the formulation of gauge theories on random lattices [5] and in close analogy construct an improved action function, which should correct some of the defects of the above described naive discretization. In a first step we project sites, links and plaquettes of the manifolds $SH(L)$ and $SH'(L)$ (dual) onto the surface of 4-dimensional spheres. We then consider an action of the form

$$S = -\sum_P \frac{A'_P}{A_P}[\beta \cos(U_P) + \gamma \cos(2U_P)], \qquad (2)$$

where each plaquette has now a weight, which is given by the ratio of its dual plaquette area divided by the plaquette area. A triangularization making use of euclidean distances approximates the plaquette areas $A_P$ and $A'_P$. Note that this lattice gauge theory behaves correctly in the naive (small coupling) continuum limit. In addition this mapping of the regular manifold $SH(L)$ onto the sphere preserves the intrinsic lattice dimensionality $D_H = 4$.

## 3. Results

We have studied the above theory (2) at two values of the parameter $\gamma = 0$ and $\gamma = -0.2$. We simulate the theory using a 3-hit Metropolis update (including overrelaxation steps at $\gamma = 0$) on a sequence of lattice sizes constructed in the above mentioned way from $SH(L)$ lattices with $L = 4, 6, 8$ for both values of $\gamma$, and $L = 10$ at $\gamma = 0$. In the vicinity of the finite volume phase transition we simulate at fixed value of $\gamma$ several $\beta$-values and employ the Ferrenberg Swendsen re-weighting method for the construction of continuous thermodynamic quantities as function of $\beta$. For each lattice size we constructed typically $(0.5 - 1) \times 10^6$ configurations. We consider moments of the plaquette energy i.e., the specific heat $C_V$ and other cumulants like the Binder cumulant $V_{BCL}$ and a quantity $U_4$ (see e.g. [4] for definitions). These quantities develop extrema at criticality and their locations define pseudocritical couplings.

Our numerical analysis indicates a critical behavior of the theory at the $\gamma$-value of $\gamma = -0.2$. Even at $\gamma = 0$ the behavior on the given lattice sizes is consistent with a critical point. To illustrate these statement we display in fig. 2 the probability distribution of the plaquette energy $P(E)$ on a $L = 8$ lattice at $\gamma = -0.2$ at the peak position of the specific heat. The variable $E$ denotes the $\beta$-term of the action (2). The distribution function is consistent with a Gaussian form, which has to be contrasted with the data of fig. 1. Such Gaussian forms are observed for all our lattice sizes at $\gamma = -0.2$, small deviations from this form show up for the data at $\gamma = 0$. In fig. 3 we display our data for the maxima of the specific heats in a double logarithmic plot as a function of the linear size scaling variable $L$. Assuming standard finite size scaling arguments at a critical point we expect $C_{V,max}$ to scale $\propto L^{\frac{\alpha}{\nu}}$. Assuming furthermore the hyperscaling relation $\alpha = 2 - D\nu$ we expect a behavior $\propto L^{\frac{2}{\nu}-D}$. The observed slopes are $\frac{\alpha}{\nu} = 1.47(20)$ and $1.18(30)$ for $\gamma = 0$ and $\gamma = -0.2$ respectively. These values are far away from the value 4 expected for a first order phase transition. We find $\nu = 0.36(2)$ for $\gamma = 0$ and $\nu = 0.38(4)$ at $\gamma = -0.2$. In addition we have checked the finite size scaling of pseudo-critical $\beta$-values, which at criticality are supposed to scale $\beta_{pc}(L) = \beta_{pc}(L = \infty) + c/L^{\frac{1}{\nu}}$. The data are consistent with such scaling laws and values for $\nu$ close to mentioned ones.

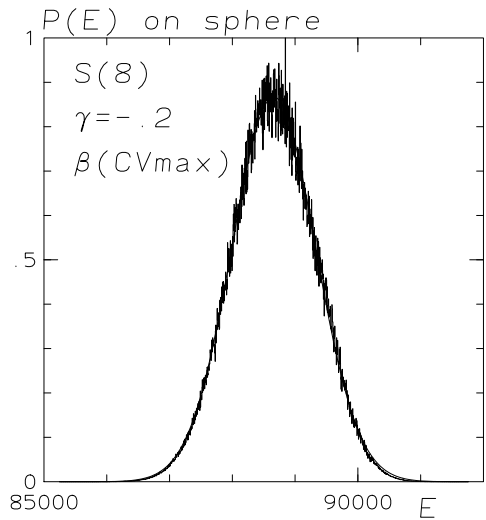

Figure 2. $P(E)$ on sphere at $\gamma = -0.2$.

## 4. Conclusion

Using a regularization of CQED on a sphere we observe a critical, continuous behavior of thermodynamic quantities for the value of $\gamma = -0.2$.

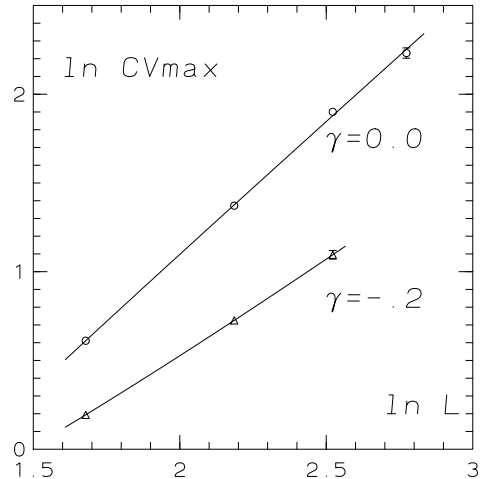

Figure 3. Maxima of the specific heat.

The observed FSS behavior is consistent with the singular behavior of a critical point, i.e. it is suggested that the theory (2) exhibits a correlation length divergence there. At $\gamma = 0$ we observe an "effective" critical behavior, but due to the limited lattice sizes we cannot draw a definite conclusion. The measured $\nu$ values indicate a nonstandard situation, where the value $\nu = \frac{1}{2}$ does not seem to be realized. This observation is connected to the question whether the underlying topological disorder transition constitutes a new universality class in $4D$ LGT. We are currently accumulating more data on larger lattices and at $\gamma = -0.5$. Further analysis, including a study of zeros of the complex partition function, then will allow us to sharpen our statements.